# Raman sensitivity to crystal structure in InAs nanowires

**Jaya Kumar Panda[1a)], Anushree Roy[1], Achintya Singha[2], Mauro Gemmi[3a)], Daniele Ercolani[4a)], Vittorio Pellegrini[4], Lucia Sorba[4]**

[1]*Department of Physics, Indian Institute of Technology Khragpur. Pin 721302. India*

[2]*Department of Physics, Bose Institute, Kolkata, Pin 700009, India*

[3]*Center for Nanotechnology Innovation @ NEST, Istituto Italiano di Tecnologia, Piazza S. Silvestro 12, I-56127 Pisa, Italy*

[4]*NEST-Istituto Nanoscienze-CNR and Scuola Normale Superiore, Piazza S. Silvestro 12, I-56127 Pisa, Italy*

**Abstract**

We report a combined electron transmission and Raman spectroscopy study of InAs nanowires. We demonstrate that the temperature dependent behavior of optical phonon energies can be used to determine the relative wurtzite fraction in the InAs nanowires. Furthermore, we propose that the interfacial strain between zincblende and wurtzite phases along the length of the wires manifests in the temperature-evolution of the phonon linewidths. From these studies, temperature-dependent Raman measurements emerge has a non-invasive method to study polytypism in such nanowires.

Arsenide based III-V semiconductors are usually found in the zinc blende (ZB) phase. However, it is now well established that, by controlling growth parameters, it is possible to obtain nanowires (NWs) of the above class of materials in wurtzite (WZ) phase as one of the main crystalline structures.[1-5] In general, high resolution transmission electron microscopy and electron diffraction measurements are used to detect crystal structures in NWs.[3,4,5,6] In addition, Raman spectroscopy and imaging have been largely used to probe the structural evolution of InAs NWs, fabricated under varying growth conditions.[1,7,8] However, no experiments to date have been able to extract the crucial information on interfacial strain between ZB and WZ phases along the length of the NW. Here we address this issue by analyzing the Raman line width of the phonon mode. In addition, we prescribe a technique to compare the relative WZ phase in NWs using temperature dependent Raman measurements. This approach offers an easy alternative to TEM measurements to compare the polytypism in NWs.

Aligned InAs NWs are grown on InAs substrate {111} using chemical beam epitaxy technique, varying the group III/V flux ratio to tune the WZ and ZB fraction of crystal phases along their length. Sample NW1082 (NW1249) has been grown at $425\pm5^{o}C$ ($430\pm5^{o}C$) with MO line pressures of 0.3 and 1.0 (2.0) Torr for TMIn and TBAs. The crystal structure of the NWs is studied using a Zeiss Libra 120 transmission electron microscope (TEM). Raman scattering measurements are carried out in back-scattering geometry using a micro-Raman spectrometer equipped with 488 nm (2.55 eV) of 5 mW air-cooled $Ar^+$ laser as the excitation light source, a spectrometer (model TRIAX550, JY) and a CCD detector. Raman spectra of all samples, over the temperature range between 120K and 230K, have been recorded using a sample cell (Model

Link-600S, JY). For TEM (Raman) measurements the NWs are mechanically transferred from the substrate to a 300 mesh copper TEM grids coated with a carbon film (silicon substrate).

Fig 1(a) and (b) show the TEM images of NW1082 and NW1249. The electron diffraction patterns from these two samples are shown in Fig. 1(c) and (d), respectively. All spots belong to WZ (in [2-1-10] zone axis) phase except the weak elongated diffuse spots (reflections belonging to ZB phase). Both possible [110] twinned orientations of ZB, which correspond to the two possible cubic stacking sequence of hexagonal array (ABC or ACB) are represented by the ZB spots that, in fact, have a mirror symmetry with respect to the $(111)_{ZB}$ reciprocal direction. At the left upper corners of Fig. 1(c) and (d) magnified parts of the patterns corresponding to the marked areas are displayed. In the diffraction pattern of NW1249 the diffused spots are clearly visible, whereas they are extremely weak for NW1082.

In order to measure the length fraction of WZ and ZB in both samples in a comparable way, we collected dark field TEM images of the NWs oriented in [2-1-10] zone axis with the objective diaphragm placed as indicated by a white circle in Fig. 1 (d). The diaphragm selects two WZ spots and one ZB spot that belongs to one of the two possible twins of the ZB structure. In this configuration the WZ and ZB structures have a strong and well-defined contrast difference and the ZB segment appears brighter or darker than WZ, depending whether it belongs to the twin individual which gives the spots in the diaphragm or to the other individual, respectively. In the upper corner of Fig. 1(a) and (b) we have shown magnified part of images of the NWs with estimated length of WZ phases. Similar measurements have been carried out for the full length of the NWs. To minimize the error in measurements, we chose five individual NWs of each sample and estimated the length of WZ and ZB phases for both the samples. The fraction ($R_{WZ}$) of WZ to the total length of the NW has been estimated in the samples NW1082 and NW1249 to be 95% and 83%, respectively. Hence, the ratio of WZ phase fraction between NW1082 and NW1249, is 1.14.

Representative Raman spectra of bulk InAs and vertically aligned as-grown InAs NWs, NW1082, in $z(xx)\bar{z}$ and $z(xy)\bar{z}$ scattering geometries and recorded at 140 K, are shown in Fig. 2(a) and (b). A recent report[8] on polarized Raman measurements on single InAs NW, substantiated by a calculation based on extended rigid ion model, provides Raman selection rules for the phonon modes of InAs NWs of mixed phase. Following the selection rules of the Raman tensor,[9,10] for bulk InAs [111] both the TO phonon at 217 cm$^{-1}$ and the LO phonon at 239 cm$^{-1}$ appear in $z(xx)\bar{z}$ configuration, whereas only TO phonon is prominent in $z(xy)\bar{z}$ configuration, in Fig. 2(a). Due to the breakdown in crystal symmetry in NWs, both LO and TO modes appear in both scattering geometries. In the observed spectra folded TO mode of the WZ phase at 212 cm$^{-1}$ and TO mode of ZB and WZ phase at 217 cm$^{-1}$ could not be resolved.

Temperature dependent Raman measurements have been carried out to compare relative strain and polytypism in NWs. Each measured spectrum was fitted with two Lorentzian functions for LO and TO modes keeping mean frequency of the TO modes ($\omega_{TO}$), LO mode ($\omega_{LO}$), widths ($\Gamma_{TO}$ and $\Gamma_{LO}$) and intensities of both LO and TO modes as free fitting parameters. As mentioned earlier, information of both ZB and WZ phases are present in the unresolved TO modes of NWs. Variation of $\omega_{TO}$ over the temperature range between 120 K and 230 K for bulk and NWs of InAs are plotted in Fig. 2(c). The magnitudes of the slopes are −0.0090, −0.0052 and −0.0057 cm$^{-1}$/K for bulk InAs, NW1082 and NW1249.

The pure anharmonic contribution on phonons, and thermal expansion of solid, changes in phonon frequency $\omega(T)$ with temperature.[11-13] The temperature derivative of $\omega(T)$ for higher temperature range (yet, below Debye temperature) is independent of temperature.[14] The variation of $\omega_{TO}$ with temperature for bulk InAs (black symbols in Fig 2(c)) must be related to anharmonicity and thermal properties of the ZB phase of InAs. As both phases are present in NW1082 and NW1249, for a particular NW sample with mixed phase, variation of $\omega_{TO}$ with temperature can be approximated as, $\frac{d\omega_{TO}}{dT} = a \times \left(\frac{d\omega}{dT}\right)_{WZ} + (1-a) \times \left(\frac{d\omega}{dT}\right)_{ZB}$, where $a$ and (1-$a$) are the fractions of WZ and ZB phases in a NW. $(d\omega/dT)_{WZ}$ and $(d\omega/dT)_{ZB}$ are the slopes of the plots related to pure WZ and ZB phases of the semiconductor. We assume that the WZ fraction in NW1249 and NW1082 is $y$ and $x$, respectively. Using the estimated values of $(d\omega/dT)_{ZB}$ (from the data for bulk InAs) and $(d\omega_{TO}/dT)$ for NW1082 and NW1249 from Fig. 2(c), the *relative fraction* (*x/y*) of the WZ phase in NW1082 and NW1249 has been estimated to be 1.13. This ratio matches excellently well with the experimental results derived from TEM measurements which is 1.14.

We now proceed to explain how the interfacial strain between two phases in InAs NWs manifests in the variation in width of the LO phonon mode ($\Gamma_{LO}$) with temperature. The changes in $\Gamma_{LO}$ with temperature for bulk and NWs of InAs, below Debye temperature (255 K), are plotted in Fig. 3(a)–(c). The anharmonicity in the vibrational deformation potential leads to the increase in $\Gamma_{LO}$ with temperature, as observed for bulk InAs (in Fig. 3(a)). For NW1082 and NW1249 the variation of $\Gamma_{LO}(T)$ with temperature is observed to be very different. Moreover, we find that the value of $\Gamma_{LO}$ at 120 K is ~5 cm$^{-1}$ for bulk InAs; while the value of $\Gamma_{LO}$ at 120K for NW1082 (6 cm$^{-1}$) is less than the same for NW1249 (8 cm$^{-1}$), however higher than the same for bulk InAs. It is well known that the width of a Raman line is strongly related to the strain in the layer.[15] The larger value of $\Gamma_{LO}$ in NWs, most likely, is the manifestation of interfacial strain between two phases along the NW. Having more number of interfaces, as seen in Fig. 1, the contribution of the interfacial strain between two phases is expected to be more in NW1249 than in NW1082. We believe that the relaxation of the strain with increase in temperature overshadows the effect of anharmonicity and leads to a negative slope of $d\Gamma_{LO}/dT$ for NW1249. It appears that the effect of anharmonicity nearly cancels the effect of relaxation of strain with temperature for NW1082. At 250K the width is observed to be same as that of bulk InAs (~ 6 cm$^{-1}$) for both NWs. Beyond 250K the variation of $\Gamma_{LO}$ with T for NWs and bulk is observed to be very similar in nature. Here we would like to mention that, in above, while analyzing the $\omega_{TO}$ mode to estimate the relative WZ fraction, the shift in one TO component due to compressive strain in one phase nearly nullifies the opposite shift in TO component due to the tensile strain in the other phase. It is to be recalled that the observed TO peak is the convolution of TO modes from WZ and ZB phases which could not be resolved. Thus, the variation of unresolved *mean* $\omega_{TO}$ with temperature is mostly dominated by the effect of anharmonicity and thermal expansion, as discussed.

In conclusion, we have demonstrated that Raman scattering measurements, like TEM, can be used as a spectroscopic tool to estimate the relative WZ phase in III-V semiconductor NWs with mixed phases. The manifestation of interfacial strain between two phases in the crystal structure is addressed from the analysis of Raman line profile.

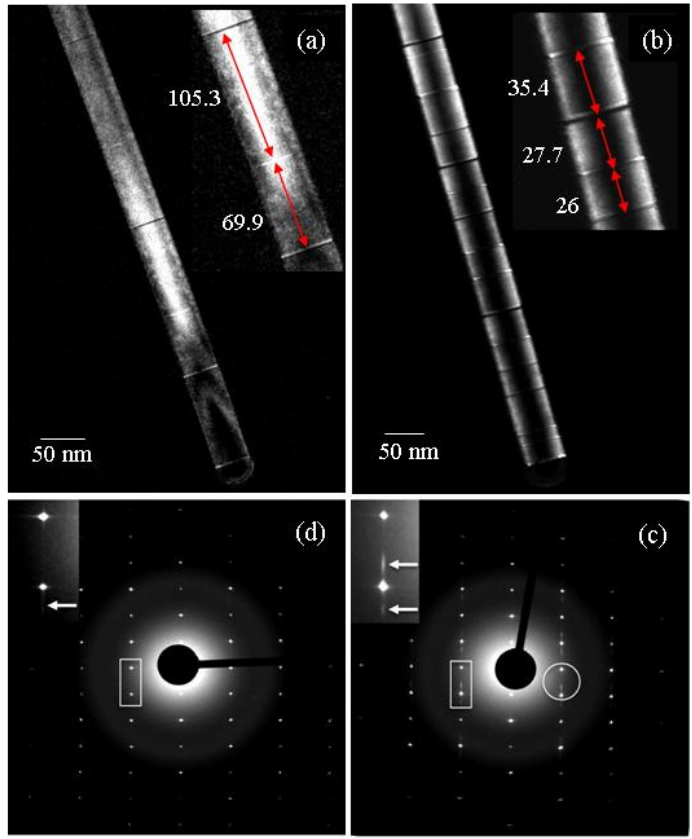

Figure. 1 TEM dark field images of InAs Nws of sample (a) NW1082 and (b) NW1249. Right hand corner of (a) and (b) show the magnified view of a part of the NW. The red arrows show the length of the WZ parts. (c) Electron diffraction patterns for sample NW1082 and (d) NW1249. Both patterns can be indexed as [2-1-10] WZ zone axes. At the left upper corners magnified parts of the patterns corresponding to the rectangular white boxes are displayed. The contrast is enhanced in order to visualize the weak spots corresponding to the ZB phase (indicated by arrows). The white circle indicates the position and the size of the objective diaphragm used for the dark field images in (a) and (b).

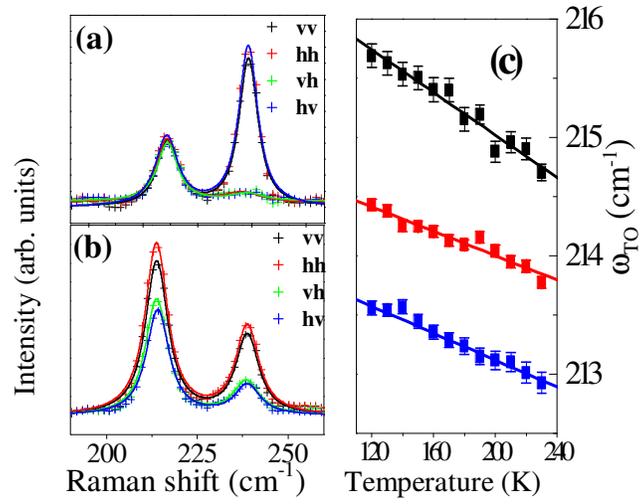

Figure 2. Raman spectra of (a) bulk InAs and (b) as-grown InAs NWs (NW1082) in parallel (VV and HH) and cross (VH and HV) polarization, recorded at 140K. (c) Variation of $\omega_{TO}$ with temperature (filled symbols). The solid lines are the best fit to the data points, from which the slopes are calculated. Color code: black for bulk InAs, red for NW1082 and blue for NW1249.

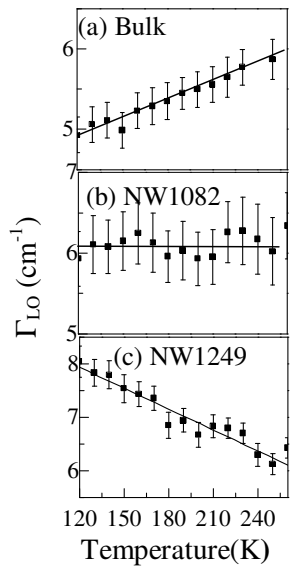

Figure 3. Variation of $\Gamma_{LO}$ with temperature for (a) bulk InAs, (b) NW1082 and (c) NW1249.

[1] I. Zardo, S. Conesa-Boj, F. Peiro, J. R. Morante, J. Arbiol, E. Uccelli, G. Abstreiter, A. Fontcuberta, I. Morral, *Phys. Rev. B* 80, 245324 (2009).

[2] F. Zhou, A. L. Moore, J. Bolinsson, A. Persson, L. Fröberg, M. T. Pettes, H. Kong, L. Rabenberg, P. Caroff, D. A. Stewart, N. Mingo, K. A. Dick, L. Samuelson, H. Linke, L.Shi, *Phys. Rev. B* 83, 205416 (2011).

[3] P. Caroff, K. A. Dick, J. Johansson, M. E. Messing, K. Deppert, L. Samuelson, Nature *Nanotechnology* 4, 50 (2008).

[4] J. Johansson, K. A. Dick, P. Caroff, M. E. Messing, J. Bolonsson, K. Deppert, L. Samuelson, *J. Phys. Chem. C* 114, 3837 (2010).

[5] D. Kriegner, C. Panse, B. Mandl, K. A. Dick, M. Keplinger, J. M. Persson, P. Caroff, D.Ercolani, L. Sorba, F. Bechstedt, J. Stangl, G. Bauer, *Nano Lett*. 11, 1483 (2011).

[6] C. Thelander, P. Caroff. S. Plissard, A.W. Dey, K.A. Dick, *Nano Lett*. 11, 2424 (2011)

[7] N. Begum, M.Piccin, F. Jabeen, G. Bais, S. Rubini, F. Martelli, A. S. Bhatti, *J. Appl. Phys*. 104,104311 (2008).

[8] N. Begum, A. S. Bhatti, F. Jabeen, S. Rubini, F. Martelli, *J. Appl. Phys*. 106,114317 (2009).

[9] M. Möller, M.M. de Lima Jr. , A. Cantarero; *Phys. Rev. B* 2011, 84, 085318 (2011).

[10] R. C. C. Leite, J. F. Scott, *Phys. Rev. Lett.* 22, 1073 (1968).

[11] T.R. Ravindran, A.K.Arora, T.A. Mary, *Phys. Rev. B* 67, 064301 (2003).

[12] H. H. Burke, I. P. Herman, *Phys. Rev. B* 48, 15016 (1993).

[13] M. Balkanski, R. F. Wallis, E.Haro, *Phys. Rev. B* 28, 1928 (1983).

[14] G.S. Doerk, C. Carraro and R. Maboudian, *Phys. Rev. B* 80, 073306 (2009).

[15] D.J. Evans, S. Ushioda, *Phys. Rev. B* 9, 1638 (1974).